\documentclass[11pt,nofootinbib,showpacs]{revtex4}
\usepackage{amsfonts,amssymb,amsmath,graphicx}
\def\dac{\displaystyle\frac}
\def\[{\left[}
\def\]{\right]}
\def\({\left(}
\def\){\right)}
\def\gammad{\gamma _{\left( \mathbf{D}\right)}}
\def\gammaterm{\gamma _{\left( \mathbf{D}\right)}+\dot b(t)^2}

\newcommand{\const}{\mathop{\rm const}\nolimits}

\begin{document}
\baselineskip7mm

\title{Dynamical compactification in Einstein-Gauss-Bonnet gravity from geometric frustration}

\author{Fabrizio Canfora}
\affiliation{Centro de Estudios Cientificos (CECS), Casilla 1469 Valdivia, Chile}
\affiliation{Universidad Andres Bello, Av. Republica 440, Santiago, Chile}
\author{Alex Giacomini}
\affiliation{Universidad Austral De Chile, Valdivia, Chile}
\author{Sergey A. Pavluchenko}
\affiliation{Instituto de Ciencias F\'isicas y Matem\'aticas, Universidad Austral de Chile, Valdivia, Chile}

\begin{abstract}
In this paper we  study dynamical compactification in  Einstein-Gauss-Bonnet gravity from  arbitrary dimension for generic values of the coupling constants.  
We showed that, when the curvature of the extra dimensional space  is negative, for any value of the spatial curvature of the four dimensional space-time one obtains a realistic behavior in which
for asymptotic time both the volume of the extra dimension and expansion rate of the four dimensional space-time tend to a constant. Remarkably, this scenario appears within the
open region of parameters space for which the theory does not admit any maximally symmetric $(4+D)$-dimensional solution, which gives to the dynamical compactification an interpretation
as geometric frustration. In particular there is no need to fine-tune the coupling constants of
the theory so that the present scenario does not violate ``naturalness hypothesis''. Moreover we showed that with increase of the number of extra dimensions the stability properties of the solution
are increased.
\end{abstract}


\maketitle

\section{Introduction}

The idea that space-time may have more than four dimensions goes back to Kaluza and \mbox{Klein~\cite{KK1, KK2, KK3}}. They introduced one extra space dimension in an attempt to unify gravity with the
electromagnetic interaction. There of course arise the question of why this extra dimension is not visible to us. A simple explanation of this fact is to assume that the extra dimension is compactified
to a very small circle. Kaluza-Klein theory can be extended to include also non-Abelian gauge fields using more extra dimensions.

Extra dimensions appear naturally in the context of string theories as well. An interesting feature is that the low energy  sector
of some string theories (see, for instance, the discussion in \cite{GastGarr}) is described by an Einstein-Gauss-Bonnet theory rather then General Relativity. Einstein-Gauss-Bonnet (EGB) theory is
actually the simplest generalization of General Relativity (GR) to
higher dimensions in the sense that it leads to second order differential equations in the metric even if the action contains higher powers of the curvature. In four dimensions the Gauss-Bonnet term
does not affect the equations of motion as it is a boundary term. However in any space-time dimension higher than four its variation gives a non-trivial contribution to the equations
of motion. Einstein-Gauss-Bonnet gravity is member of a larger family of higher curvature gravity theories: indeed  in any higher odd dimension it is possible to add to the gravitational action
another higher power curvature term (in five dimensions -- quadratic, in seven -- cubic, in nine -- quartic and so on) in such a way that the resulting equations of motion remain of second order. This family of
gravity theories is called Lovelock gravity~\cite{Lovelock}.

One peculiar feature of the Einstein-Gauss-Bonnet theory, which distinguish
it from standard General Relativity, is that the field equations do not
necessarily imply the vanishing of torsion \cite{TZ-CQG}, so that in the
first order formalism, besides the graviton, there are additional
propagating degrees of freedom related to the torsion.
Indeed  exact solutions with non-trivial torsion have been found \cite{CGT07, CGW07, CG, CG2, ACGO}, but here we will consider the torsion free case.

The most generic Einstein-Gauss-Bonnet action in any space-time dimension higher than four is the sum of three terms namely an Einstein-Hilbert term $\int \sqrt{-g} R$, a
Gauss-Bonnet term $\int\sqrt{-g}(R_{\alpha\beta\mu\nu}R^{\alpha\beta\mu\nu}-4R_{\mu\nu}R^{\mu\nu}+R^2)$ and a volume term $\int \sqrt{-g}\Lambda$. The volume term in the context of General Relativity is
called the ``cosmological term'' however, as we will see in the next section, in the presence of a Gauss-Bonnet term, the value of the effective cosmological constant is actually a function of all three
coupling constants and not just of the volume term. Indeed this theory being quadratic in the curvature can have up to two different maximally symmetric constant curvature solutions. The values of the
curvature of the maximally symmetric solutions are the effective cosmological constants of the theory.
Interestingly enough  it can also happen for certain regions of the parameter space that the theory does not admit any maximally symmetric
solution. This sector of the theory has not been studied very much up to now especially in the  context of cosmology but it is actually of great theoretical interest since it generates in a natural way
a symmetry breaking mechanism induced by geometry in which there is no solution with the maximum number of Killing fields which are allowed in principle by the geometry. In this sense one could speak
of ``geometric frustration''\footnote{In statistical mechanics of disordered systems (see, for two classic reviews, \cite{glass1} and \cite{glass2}) the terms ``frustration'' refers generically to 
situations in which the non-trivial geometry of the graph on which the spin Hamiltonian is defined prevents the system itself from reaching a state in which the interactions energies of all the pairs 
of neighboring spins have the minimal value. In a standard ferro-magnetic systems one can always reach a state in which the interactions energies of all the pairs of neighboring spins have the 
minimal value (it is enough to put all the spin variables in the same state). On the other hand, there are many systems (such as spin glasses) in which for topological reasons it is not possible to 
satisfy all the constrains to minimize the pairwise interactions energies. In particular, the same spin-Hamiltonian may or may not have frustration depending on the graph on which it is analyzed. 
Hence, in the present case we use the term ``frustration'' in analogy with statistical mechanics since, in an open region of the parameters space, the maximally symmetric configuration (which is in 
principle allowed by the geometry) is not a solution of the theory.}.

In  cosmology the addition of Gauss-Bonnet term to the action is of special interest in order to study how it affects the size evolution of the extra dimensions
(see \cite{add_1, add_2, add_3, add_4, add_5, add_6, add_7, add_8, add_9, add_10, add13, add_11, add_12, mpla09, prd09, grg10, gc10, prd10, new12, Germani:2002pt, Kofinas:2003rz, Papantonopoulos:2004jd, 44,
BD85, Is86, MO04} for older and recent developments in cosmology with
Gauss-Bonnet and more general Lovelock terms; for an updated review and references see \cite{1.5}). In the context of extra dimensions
the most important question is why the extra dimensions are small and of approximately constant size while the three space dimensions are much larger and expanding. The extra dimensions may
have been much larger in the far past and non-constant in size. It is therefore of great interest to find a dynamical mechanism of compactification dictated just by the equations of motion derived from
the gravitational action. A sensible requirement is that such a mechanism should not involve any fine-tuning of the coupling constants of the theory. Indeed, if the compactification mechanism only works
for a precise value of the couplings of the theory, any small change could destroy it. Moreover in cosmology ``naturalness'' is often advocated, namely good phenomenological behavior should be obtained
without assuming that one of the couplings of the theory is much smaller than the others.
Such a dynamical mechanics was proposed in \cite{add13} for (5+1)-dimensional EGB theory; but we are proposing here a more generic setup with arbitrary number of extra dimensions, curvature in 
both manifolds and a volume term (Lambda term) in the action which opens new scenarios.

In \cite{CGTW}, it was shown for the first time  how to construct a
realistic static compactification (namely, the extra dimensions do not evolve in time) in seven (or higher) dimensions in Lovelock
gravities. A suitable class of cubic Lovelock theories allows to recover
General Relativity with a small cosmological constant in four dimensions
with the extra dimensions of constant curvature. However, in the cases analyzed in \cite{CGTW}, in order to get the desired compactification both a fine-tuning (namely, one of the couplings is a
function of the others) and a violation of ``naturalness'' are necessary.
Here we want to improve such situation in a cosmological context within the framework of Einstein-Gauss-Bonnet theory.

We will therefore use the ansatz of a space-time which is a warped product of the form \mbox{$M_4\times b(t)M_D$}, where $M_4$ is a Friedmann-Robertson-Walker manifold with scale factor $a(t)$ whereas
$M_D$ is a D-dimensional Euclidean compact and constant curvature manifold with scale factor $b(t)$ and study the evolution of the two scale factors. Due to the highly nonlinear nature of the
equations of motion it is not possible to integrate them in a closed form. However it is possible to understand in detail all the relevant features of the theory depending on the values of the couplings
and of the curvature of space and extra dimension by performing a numerical analysis.

In most cases, when considering the Gauss-Bonnet, or even higher-order Lovelock gravity, in literature  spatially flat sections are considered. In our paper we decided to
consider also the case with non-zero constant curvature as it allows us to
see the influence of the curvature on the cosmological dynamics. Despite the fact that, according to current observational cosmological data, our Universe is flat with a high precision, at the early
stages of the Universe evolution the curvature could comes into play. In there, negative curvature only ``helps'' inflation (since the effective equation of state for negative curvature is $w=-1/3$),
while the positive curvature affects the inflationary asymptotics, but its influence is not strong for a wide variety of the scalar field potentials (see \cite{infl1, infl2} for details), so that
we can safely consider both signs for curvature without worrying for inflationary asymptotics.
%We will however see that if we insist on a late time evolution  compatible with experimental data we must put spatial curvature of the $M_4$ part to zero.

The structure of the paper is the following: in the second section we will give a basic review on the relevant features of Einstein-Gauss-Bonnet theory especially the relation between the couplings of
the theory and the effective cosmological and Newton constant.
In the third section we will write down the equations of motion for our metric ansatz in generic dimension and for generic space and extra dimension curvature.
Later in the same section we will present the results, as well as discuss the stability of the solution in the large number of extra dimensions. In the last section the conclusions will be drawn.

\section{The Einstein-Gauss-Bonnet in arbitrary dimension}
The Einstein-Gauss-Bonnet action in arbitrary dimension possesses 3 coupling constants and in the vielbein formalism reads

\begin{equation}
S=\int \epsilon_{{A_1} \ldots A_{D+4}}(\frac{c_0}{D+4}e^{A_1}\ldots e^{A_{D+4}}+\frac{c_1}{D+2}R^{{A_1}{A_2}}e^{A_3}\ldots e^{A_{D+4}}+\frac{c_2}{D}R^{A_1 A_2}R^{A_3 A_4}e^{A_5}\ldots e^{A_{D+4}})
\label{actionEGB}
\end{equation}

The couplings $c_2$, $c_1$ and $c_0$ correspond the the Gauss-Bonnet term, Einstein-Hilbert term and the volume ``cosmological'' term respectively.
When the Gauss-Bonnet coupling $c_2$ is zero the coupling $c_0$  is just the ``cosmological constant'' and the coupling $c_1$ is  just be the Newton constant. However if $c_2$ is non-vanishing
this is in general not true. Indeed the cosmological constant is the quantity which gives the curvature of the maximally symmetric solution of the theory. It is therefore useful to study the equations
of motion obtained by varying the action with respect to the vielbein (which in the metric formalism correspond to the Einstein-Gauss-Bonnet field equations), they read

\begin{equation}
\epsilon_{A_1}=\epsilon_{{A_1} \ldots A_{D+4}}({c_0}e^{A_2}\ldots e^{A_{D+4}}+{c_1}R^{{A_2}{A_3}}e^{A_4}\ldots e^{A_{D+4}}+{c_2}R^{A_2 A_3}R^{A_4 A_5}e^{A_6}\ldots e^{A_{D+4}})\label{EOMD+4}.
\end{equation}

The value of the $(4+D)$-dimensional ``cosmological constant'' is given by  the value of the constant curvature of the maximally symmetric space-time solutions. The ansatz for a constant curvature space-time
in the vielbein formalism reads

\begin{equation}
R^{AB}=\Lambda_{D+4}e^{A}e^{B}\label{constantcurv},
\end{equation}
where of course $\Lambda_{D+4}$ must be real.
Plugging this ansatz into the equations of motion one gets a polynomial in $\Lambda_{D+4}$

\begin{equation}
(c_2\Lambda_{D+4}^2 + c_1\Lambda_{D+4}+c_0)e^{A_1}\ldots e^{A_{D+4}}=0\label{polynomial};
\end{equation}

\noindent this equation admits as solution

\begin{equation}
\Lambda_{eff}\equiv\Lambda_{D+4}= \frac{-c_1\pm \sqrt{(c_1)^2-4c_2c_0}}{2c_2}\label{Leff}.
\end{equation}

This implies that in general the cosmological constant
is a function of all three couplings of the action and not just $c_0$. If the discriminant is positive, the theory possesses  two different possible cosmological constants which can even have opposite
sign. A special case exists when the couplings are fine tuned in such a way that the discriminant is zero then the two roots of the polynomial are degenerate. In the case that $c_0\neq 0$ the
discriminant can also be negative. In this case the two possible values for the cosmological constant are complex which means that in this case no maximally symmetric solution exists at all. This
interesting phenomenon can be interpreted as induced by ``geometric frustration'' which prevents the existence of metrics preserving all the symmetries which in principle are available. It is worth noting that
this can only happen when the highest power in the curvature of the Lovelock action is even (otherwise at least one real root would always exist). Moreover, the parameters region in which (at least) 
a maximally symmetric vacuum exist has been already extensively analyzed in the literature. Since it is known that, in such a region, it is not possible to obtain a dynamical compactification which 
is both free of fine-tunings and free of violations of the naturalness hypothesis, we will focus on the region in which geometric frustration occurs.

Another important feature of this theory, in opposition to GR, is that by compactifying the space-time to $M_4\times M_D$ where $M_4$ is a four dimensional space-time and $M_D$ is some compact manifold with
constant curvature $\Lambda_D$ is that the Newton constant of the effective four dimensional theory is not just proportional to $c_1$. This can be seen by projecting the $4+D$ dimensional equations down to four dimensions

\begin{equation}
\begin{array}{l}
\epsilon_{i}=\(c_1(D+1)+2c_2(D-1)\Lambda_D\)R^{jk}+\(c_0\frac{(D+3)(D+2)(D+1)}{6}+\right. \\ \left.
+c_1\Lambda_D\frac{(D+1)D(D-1)}{6}+c_2\Lambda_D^2\frac{(D-1)(D-2)(D-3)}{6}\)e^ie^j=0\label{effectivenewton}.
\end{array}
\end{equation}
where the lowercase indices run from zero to three.
The term $(c_1(D+1)+2c_2(D-1)\Lambda_D)$ which multiplies the four dimensional curvature two form $R^{jk}$ is the ``effective Newton constant'' whereas
the term $(c_0\frac{(D+3)(D+2)(D+1)}{6}+c_1\Lambda_D\frac{(D+1)D(D-1)}{6}+c_2\Lambda_D^2\frac{(D-1)(D-2)(D-3)}{6})$ is an ``effective 4-dimensional cosmological constant''.
This means that if the Gauss-Bonnet term does not vanish and moreover the $D$-dimensional curvature does not vanish the effective Newton constant is not just proportional to $c_1$. In particular, the
effective Newton constant  can even have a negative sign.

\section{Dynamical compactification}

Nowadays it is widely accepted  that the
search for a unified theory  requires additional space-time
dimensions, the EGB theory being the simplest generalization of General Relativity to higher dimensions.

In the present paper we want to show that in a cosmological context one can get a dynamical compactification in an open region of the parameter space
(and in agreement with the ``naturalness hypothesis'') in which a potentially $4+D$ universe evolves towards a $4\times D$ universe in which the size of the extra dimensions is much smaller than the
size of the 4 macroscopic dimensions. The key feature of EGB theory which gives rise to  a phenomenologically realistic dynamical compactification is the possibility of geometric symmetry
breaking mentioned above which can only occur in Lovelock theories in which the highest power of the curvature tensor is even.

The ansatz for the metric is
\begin{equation}
ds^{2}=-dt^{2}+a(t)^{2}d\Sigma _{(3)}^{2}+b(t)^{2}d\Sigma _{(\mathbf{D}%
)}^{2}\ ,  \label{Ansatz-metric}
\end{equation}%

\noindent where $d\Sigma _{(3)}^{2}$ and $d\Sigma _{(\mathbf{D})}^{2}$ stand for the
metric of two constant curvature manifolds $\Sigma _{(3)}$ and $\Sigma
_{(\mathbf{D})}$. It is worth to point out that even a negative constant curvature space can be compactified by making the quotient of the space by a freely acting discrete subgroup of $O(D,1)$ \cite{wolf}. The vielbein can then be chosen as
\begin{equation}
e^{0}=dt;\;\;\;e^{i}=a(t)\tilde{e}^{i};\;\;\;e^{a}=b(t)\tilde{e}^{a},
\label{ansatz}
\end{equation}%

\noindent where $\tilde{e}^{i}$ stands for the intrinsic vielbein of $\Sigma _{(3)}$, $%
\tilde{e}^{a}$ stands for the intrinsic vielbein of $\Sigma _{(\mathbf{D}%
)}^{2}$, and $\epsilon ^{ijk}$\ is the Levi-Civita symbols on $\Sigma _{(3)}$.

The spin connection  reads

\begin{eqnarray}
\omega _{\ 0}^{i} &=&\frac{\overset{.}{a}}{a}e^{i};\;\;\;\omega _{\
0}^{a}=\frac{\overset{.}{b}}{b}e^{a};\;\;\;\omega ^{ij}=\tilde{\omega}%
^{ij},  \notag \\
\omega ^{ab} &=&\tilde{\omega}^{ab};\ \ \omega ^{jb}=0  \label{levicivita}
\end{eqnarray}%
where $\tilde{\omega}^{ij}$ corresponds to the intrinsic Levi-Civita
connection of $\Sigma _{(3)}$, $\tilde{\omega}^{ab}$ corresponds to the
intrinsic Levi-Civita connection of $\Sigma _{(\mathbf{D})}$.

The Riemannian curvature reads
\begin{gather}
R^{0i}=\frac{\overset{..}{a}}{a}e^{0}e^{i}\;;\;\;\;R%
^{0a}=\frac{\overset{..}{b}}{b}e^{0}e^{a};\ \ \ \ R^{ia}=\frac{%
\overset{.}{a}}{a}\frac{\overset{.}{b}}{b}e^{i}e^{a}  \notag \\
R^{ij}=\tilde{R}^{ij}+\left( \frac{\overset{.}{a}}{a}\right)
^{2}e^{i}e^{j}\;;\;\;\;R^{ab}=\tilde{R}^{ab}+\left( \frac{\overset%
{.}{b}}{b}\right) ^{2}e^{a}e^{b}  \label{riemann}
\end{gather}%
where $\tilde{R}^{ij}$ stands for the intrinsic Riemannian curvature of the
manifold $\Sigma _{(3)}$, $\tilde{R}^{ab}$ stands for the intrinsic
Riemannian curvature of the manifold $\Sigma _{(\mathbf{D})}$ (which will
play the role of the extra-dimensional manifold). We will assume (as it is
usual in  literature) that $\Sigma _{(3)}$ and $\Sigma _{(%
\mathbf{D})}$ have constant Riemmanian curvatures given by $\gamma _{\left(
3\right) }$ and $\gamma _{\left( \mathbf{D}\right) }$. It follows that%
\begin{eqnarray*}
\tilde{R}^{ij} &=&\gamma _{\left( 3\right) }\tilde{e}^{i}\tilde{e}^{j};\ \ \
\ \ \ \;\;\tilde{R}^{ab}=\gamma _{\left( \mathbf{D}\right) }\tilde{e}^{a}\tilde{e}^{b}
\\
R^{ij} &=&\frac{\left[ \gamma _{\left( 3\right) }+\left( \overset{%
.}{a}\right) ^{2}\right] }{a^{2}}e^{i}e^{j};\;\;\;\;\;\;R^{ab}=%
\frac{\left[ \gamma _{\left( \mathbf{D}\right) }+\left( \overset{.}{b}\right) ^{2}%
\right] }{b^{2}}e^{a}e^{b}
\end{eqnarray*}

It will be convenient to use the following notation%

\begin{gather}
R^{0i}=A_{(1)}e^{0}e^{i}\;\;\;;\;\;%
\;R^{0a}=B_{(1)}e^{0}e^{a}\;;\;\;\;R^{ia}=\mathring{R}^{ia}=Ce^{i}e^{a}\ ,
\label{nota1} \\
R^{ij}=A_{(2)}e^{i}e^{j},\ \ R^{ab}=B_{(2)}e^{a}e^{b}\ ,  \label{nota2}
\end{gather}%

with%
\begin{eqnarray}
A_{(1)} &=&\frac{\overset{..}{a}}{a},\ \ \ C=\frac{\overset{.}{a}\overset{.}{%
b}}{ab},\ \ \ B_{(1)}=\frac{\overset{..}{b}}{b}, \notag \\
A_{(2)} &=&\frac{\left[ \gamma _{\left( 3\right) }+\left( \overset{.}{a}%
\right) ^{2}\right] }{a^{2}},\
\ \ B_{(2)}=\frac{\left[ \gamma _{\left( \mathbf{D}\right) }+\left( \overset{.}{b}%
\right) ^{2}\right] }{b^{2}}.\label{ABdef}
\end{eqnarray}

It will be convenient to define the following
rescaled coupling constants appearing in the original action in (\ref{actionEGB})%

\begin{equation}
\alpha =\frac{\left(D+3\right) \left(D+2\right) \left(D+1\right) }{6}c_{0}\ ,\ \ \ \beta =\frac{\left(D+1\right) D\left(D-1\right) }{6}c_{1}\ ,\ \ \ \gamma =\frac{\left(D-1\right) \left(D-2\right)
\left(D-3\right) }{6}c_{2}\ . \label{abc}
\end{equation}%

Thus, $\beta $ is related to the coupling constant of the Einstein-Hilbert
term, $\gamma $ corresponds to the Gauss-Bonnet coupling constant while $%
\alpha $ is a cosmological term. According to idea of ``naturalness''
discussed above, no one of the original
coupling constants of the theory $c_{0}$, $c_{1}$ and $c_{2}$ is privileged
with respect to the others: the three-dimensional parameters space of the
theory will be analyzed without any fine-tuning.

The structure of the torsion-free equations of motions is the following:%
\begin{equation}
\mathcal{E}_{A_{1}}=\left( c_{0}e^{A_{2}}\dots e^{A_{\mathbf{D}%
+4}}+c_{1}R^{A_{2}A_{3}}e^{A_{4}}\dots e^{A_{\mathbf{D}%
+4}}+c_{2}R^{A_{2}A_{3}}R^{A_{4}A_{5}}e^{A_{6}}\dots e^{A_{\mathbf{D}+4}}\right)
\epsilon_{A_{1}A_{2}\dots A_{\mathbf{D}+4}} =0  \label{eql}
%\epsilon_{A_{1}A_{2}\dots A_{e^{A_{6}} \dots e^{A_{\mathbf{D}+4}}}}=0  \label{eql}
\end{equation}%

Thus, the equations $\mathcal{E}_{0}=0$ and $\mathcal{E}_{i}=0$ read:%
\begin{eqnarray}
\mathcal{E}_{0} &=&0\Leftrightarrow 0=\alpha +\beta \left( B_{(2)}+\frac{6}{%
D-1}C+\frac{6}{D\left(D-1\right) }A_{(2)}\right)
+\gamma \left(  B_{(2)}^{2}+\frac{12A_{(2)}B_{(2)}}{\left(
D-2\right) \left(D-3\right) }+\right.   \notag \\
&&\left. +\frac{24C^{2}}{\left(D-2\right) \left(D
-3\right) }+\frac{12B_{(2)}C}{\left(D-3\right) }+\frac{24A_{(2)}C}{%
\left(D-1\right) \left(D-2\right) \left(D
-3\right) }\right) ,  \label{eq0}
\end{eqnarray}%
\begin{equation*}
\mathcal{E}_{i}=0\Leftrightarrow 0=\alpha +\beta \left( B_{(2)}+\frac{
4A_{(1)}}{D\left(D-1\right) }+\frac{2B_{(1)}}{D-1
}+\frac{2A_{(2)}}{D\left(D-1\right) }+\frac{4C}{\left(
D-1\right) }\right) +\gamma \left(  B_{(2)}^{2}+\right.
\end{equation*}%
\begin{equation*}
+\frac{16A_{(1)}C}{\left(D-1\right) \left(D-2\right)
\left(D-3\right) }+\frac{8B_{(2)}C}{D-3}++\frac{
8A_{(1)}B_{(2)}}{\left(D-2\right) \left(D-3\right) }+
\frac{8A_{(2)}B_{(1)}}{\left(D-1\right) \left(D-2\right)
\left(D-3\right) }+
\end{equation*}%
\begin{equation}
\left. +\frac{16B_{(1)}C}{\left(D-2\right) \left(D
-3\right) }+\frac{4B_{(1)}B_{(2)}}{\left(D-3\right) }+\frac{
4A_{(2)}B_{(2)}}{\left(D-2\right) \left(D-3\right) }+
\frac{8C^{2}}{\left(D-2\right) \left(D-3\right) }\right)
\ \ ,  \label{eqq1}
\end{equation}%
while the equation $\mathcal{E}_{a}=0$ reads%
\begin{equation*}
\mathcal{E}_{a}=0\Leftrightarrow 0=\frac{D}{\left(D-
4\right) }\alpha +\frac{\left(D-2\right) }{\left(D-
4\right) }\beta \left( B_{(2)}+\frac{6A_{(1)}}{\left(D-1\right)
\left(D-2\right) }+\frac{2B_{(1)}}{D-2}+\frac{6A_{(2)}}{%
\left(D-1\right) \left(D-2\right) }+\frac{6C}{\left(
D-2\right) }\right) +
\end{equation*}%
\begin{equation*}
+\gamma \left( B_{(2)}^{2}+\frac{48A_{(1)}C}{\left(D
-2\right) \left(D-3\right) \left(D-4\right) }+\frac{
12B_{(2)}C}{D-4}+\frac{24C^{2}}{\left(D-3\right) \left(
D-4\right) }+\right.
\end{equation*}%
\begin{equation*}
+\frac{12A_{(1)}B_{(2)}}{\left(D-3\right) \left(D
-4\right) }+\frac{24A_{(2)}B_{(1)}}{\left(D-2\right) \left(
D-3\right) \left(D-4\right) }+\frac{24B_{(1)}C}{\left(
D-3\right) \left(D-4\right) }+\frac{4B_{(1)}B_{(2)}}{%
\left(D-4\right) }+
\end{equation*}%
\begin{equation}
\left. +\frac{12A_{(2)}B_{(2)}}{\left(D-3\right) \left(D
-4\right) }+\frac{24A_{(2)}C}{\left(D-2\right) \left(D-
3\right) \left(D-4\right) }+\frac{24A_{(1)}A_{(2)}}{\left(
D-1\right) \left(D-2\right) \left(D-3\right) \left(
D-4\right) }\right)   \label{eqq2}
\end{equation}%

This notation in which the $D$ factors appear in the denominators is suitable for a large $D$ analysis in which one assumes that the physical quantities (the two scale factors in the present case) are 
analytic functions of $1/D$. Consequently, one can expand both scale factors in series of $1/D$ with time-dependent coefficients and replace the expressions into Eqs. (\ref{eq0}), (\ref{eqq1})
and (\ref{eqq2}). This allows to 
derive a set of recursive equations order by order in the $1/D$ expansion. Although the expressions of the recursive equations become quite involved after few iterations, this provides one with a 
systematic way to compute $1/D$ corrections.

 The cosmological equations are simply obtained by placing the  expressions from (\ref{ABdef}) into Eqs. (\ref{eq0}),
(\ref{eqq1}) and (\ref{eqq2}). It is worth noting here that of the three equations (\ref{eq0}), (\ref{eqq1})
and (\ref{eqq2}) only two are independent due to the Bianchi identities.

Thus, one has two evolution equations for the scale factors $a(t)$ and $b(t)$.
 Obviously, the ideal scenario is the one in which, without neither fine tunings nor violations of ``naturalness'', one gets
\begin{equation}
\begin{array}{l}
b(t)\underset{t\rightarrow \infty }{\rightarrow }b_c=\const,\\
\frac{\dot{a(t)}}{a(t)}\underset{t\rightarrow \infty }{\rightarrow }H_a=\const,
\end{array}\label{good_beh}
\end{equation}%

\noindent since in this way the theory would approach for late time to an effective four dimensional universe with small compact extra dimensions.

We leave the complete description of all possible regimes and their abundance to a separate paper, here we state that this scenario is present in a case with negative discriminant of (\ref{polynomial})
(so that $\Lambda_{eff}$ (\ref{Leff}) is imaginary)
and negative curvature of the extra dimensions ($\gammad = -1$ regardless of the $\gamma_{(3)}$).

To the best of authors knowledge this is the first time that a dynamical compactification in which the size of extra dimensions approaches to a constant while the three macroscopic extra dimensions expand has been achieved without fine tunings or ad hoc matter fields.

Finally, let us describe the region on the $\{c_0,\,c_1,\,c_2\}$ parameters space which satisfy both the condition for $\Lambda_{eff}$ to be imaginary and the condition for
effective Newton constant $G_{N,\, eff} = c_1 (D+1) + 2c_2(D-1)\Lambda_D$ to be positive. So the system of inequalities is (we put $\Lambda_D = -1$ since only this choice gives us ``well-behaved''
regime)

\begin{equation}
\begin{array}{l}
G_{N,\, eff} = c_1 (D+1) - 2c_2(D-1) > 0, \\
c_1^2 - 4c_0c_2 < 0;
\end{array}\label{region1}
\end{equation}

\begin{figure}
\includegraphics[width=1.0\textwidth, angle=0]{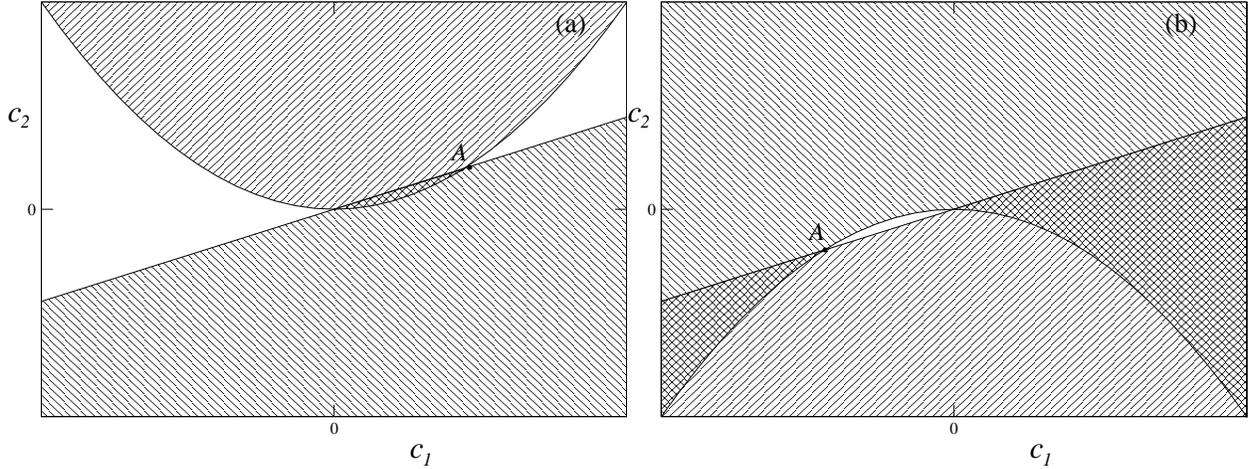}
\caption{The area on the $(c_1,\,c_2)$ space which corresponds to the requirements of the positivity of the effective Newton constant and negativity of the discriminant of (\ref{polynomial}) is
represented as double-dashed region. The (a) panel corresponds to the $c_0 > 0$ case while (b) panel -- to the $c_0 < 0$ (see text for details).}\label{region}
\end{figure}

\noindent we plot these two inequalities in the $(c_1,\,c_2)$ coordinates in Fig. \ref{region}. There we dashed with different inclinations each of (\ref{region1}) conditions so that double-dashed
region corresponds to the case where both of them satisfied. In (a) panel we presented the data for $c_0 > 0$ case while in the (b) panel -- for $c_0 < 0$. The coordinates for the $A$ point are:
$c_1 = 2c_0(D+1)/(D-1)$, $c_2 = c_0 (D+1)^2/(D-1)^2$ -- it is true for both panels since the sign for $c_0$ is different in them. So that one can see that in $c_0 > 0$ case the area for
possible $(c_1,\,c_2)$ is compact while in $c_0 < 0$ it is noncompact. Let us note also that ``classical value'' for $c_0$ is $c_0 = -2\Lambda$, where $\Lambda$ is ``classical'' cosmological constant,
so that the negative value for $c_0$ is favored both by noncompactness of the corresponding region and by ``classical argumentation''.

\subsection{Late time behavior}

In this subsection let us analyze the late-time behavior of the system. Here we apply the late-time asymptotic of the region which we are interested in 
($b(t) = b_c$, $H(t) = H_a$). Before giving the equations themselves, let us note one thing -- there are two main late-time regimes from the extra dimensions
perspective -- one with $b(t) = b_0$ and so $H_b \to 0$ and the other is $H_b > 0$ and so $\gammaterm \to H_b^2$. Obviously, the latter of them is just flat case, while former
involves curvature. The same is true for $M_4$ as well -- if we impose nonzero curvature on it, the only difference would be appearance of the regime with $a(t) = a_c = \const$ with
$H_a \to 0$, which obviously contradicts cosmological data.
So, with $b(t) = b_c$, $H(t) = H_a$ ansatz, equations (\ref{eq0})--(\ref{eqq2}) take a form:

\begin{equation}
\begin{array}{l}
\alpha + \beta y + \dac{6\beta x}{D(D-1)} + \gamma y^2 + \dac{12\gamma xy}{(D-2)(D-3)} = 0,
\end{array}\label{late1}
\end{equation}

\begin{equation}
\begin{array}{l}
\alpha D + \beta (D-2) y + \dac{12\beta x}{D-1} + \gamma (D-4) y^2 + \dac{24\gamma xy}{D-3} + \dac{24\gamma x^2}{(D-1)(D-2)(D-3)} = 0,
\end{array}\label{late2}
\end{equation}

\noindent where we  put $y = \gammad/b_c^2$ and $x=H_a^2$. It appears that (\ref{late1}) corresponds to both $\mathcal{E}_{0}=0$ and
$\mathcal{E}_{i}=0$ -- rather, we have $\mathcal{E}_{i}=\mathcal{E}_{0}$ since there is no dynamics in extra dimensions; (\ref{late2}) corresponds to $\mathcal{E}_{a}=0$. System
(\ref{late1})--(\ref{late2}) could be further simplified to one fourth-order equation with respect to $y$, but cannot be generally solved. Numerically we verified that
for wide range of  $\{c_0, c_1, c_2\}$ that gives imaginary $\Lambda_{eff}$ there are solutions of (\ref{late1})--(\ref{late2}) with $x>0$ (it must holds since $x=H_0^2$) and
only $y<0$, which gives  $\gammad=-1$.

\subsection{Stability of the solution}

In this subsection we want to check the stability of the solution $b(t) = b_c$, $H(t) = H_a$. So we perturbed it as $b(t) = b_c + \delta b$, $H(t) = H_a + \delta H$, and substitute
them into (\ref{eq0})--(\ref{eqq2}). The resulting system could be brought to one second-order equation with respect to $\delta b$:

\begin{equation}
\begin{array}{l}
a_2 \delta \ddot b + a_1 \delta \dot b + a_0 \delta b=0,
\end{array}\label{stab}
\end{equation}

\noindent where $a_i = a_i (c_0, c_1, c_2, D, \gammad)$ are very nasty polynomial coefficients. Due to the nature of the coefficients, it is almost impossible to study them
analytically, so we solved (\ref{stab}) numerically. Our numerical study revealed that the solution is stable for a wide range of the parameters and the initial conditions which
lead the $b(t) = b_c$, $H(t) = H_a$ solution.

But in the large $D$ limit (\ref{stab}) behave in a interesting way: we have $\{a_2,\, a_1\} \propto D^5$ while $a_0\propto D^6$ which implies that with increase of $D$ the solution becomes more
stable. One can make a mechanical analogy with excitation of a massive string -- increasing mass of the string makes it more difficult to excite it.
The large $D$ limit General Relativity  in the context of compactification has been also studied in \cite{CGZ, Baskal, Guendelman}. The large $D$ limit in the context of Lovelock gravity has been
studied in \cite{Giribet}. In both cases it is manifest that the large $D$ limit improves the behavior of the theory. Here we have confirmed that in this limit the stability of the dynamical
compactification regime improves as well.

\section{Discussion and comments}

\label{Discussion}

In this paper it has been shown that EGB gravity gives rise to dynamical compactification in which an initially $4+D$ Universe evolves towards a $4\times D$ Universe in which the size of the extra
dimensions are much smaller than the size of the non-compact ones. We have investigated numerically the system under consideration and found that a phenomenologically
realistic scenario happens for open region of the couplings space $\{c_0,\,c_1,\,c_2\}$  where the theory does not admit a maximally symmetric vacuum solution. This scenario
can therefore be interpreted as symmetry breaking mechanism.

Remarkably this scenario does not require neither fine-tunings nor violations of ``naturalness hypothesis''.
Independently of the sign of the curvature of the three dimensional space section the curvature of the extra dimensional space must be negative.

As we mentioned in the Introduction, the similar result -- the dynamical compactification without violation of ``naturaless'' -- was proposed in \cite{add13} for the (5+1)-dimensional
EGB theory. As one can see, our setup is different from the one used in \cite{add13} -- we considered both $M_D$ and $M_4$ as manifolds with constant and possibly nonzero curvature, which gives a rise
to additional curvature terms and, as a consequence, to a new regime. Additionally, we considered all possible geometrical terms, including the boundary term ($c_0\ne 0$), which also affects
the dynamics. Overall, despite the fact that in both cases -- in \cite{add13} and in our paper -- we can see dynamical compactification, it is brought by different phenomena. In our case it is 
geometric frustration, which is brought by a combination of nonzero curvature and nonzero boundary term -- both of them are usually omitted from consideration since they complicate the equations a lot.
But one can see that if you consider such thems, new beautiful regimes could appear; in that sence our paper holds some methodological character as well.

In the analysis of the dynamical compactification we have supposed  the torsion to be zero. However in first order formalism the equations of motion of EGB gravity do not imply the vanishing of torsion 
which is therefore a propagating degree of freedom. To study its effects in the context of dynamical compactification will be object of future investigation.

EGB gravity is the simplest generalization of general relativity within the class of Lovelock gravity. As we considered an arbitrary number of extra dimensions it would also be interesting to study 
the effect of higher order Lovelock terms in the compactification mechanism. The results obtained in this paper suggest that the effects of higher terms depend sensibly on the fact if the highest 
curvature power is even as only in this case there exist a region in the parameter space admits no maximally symmetric solution.

\bigskip

\textit{Acknowledgments.-- }
This work
was supported by Fondecyt grants 1120352, 1110167, and 3130599. The Centro de Estudios Cientificos (CECs) is funded
by the Chilean Government through the Centers of Excellence
Base Financing Program of Conicyt.
S.A.P. was partially supported by RFBR grant No. 11-02-00643.

\end{document}